\documentclass{article}

\usepackage{emulateapj,apjfonts}

\lefthead{Kifonidis et al.}
\righthead{Nucleosynthesis and Clump Formation in a Core Collapse Supernova}

\usepackage{natbib}
\usepackage{epsfig}

\newcommand{\Msol}{\rm{M_{\odot}}}
\newcommand{\Ni}{\rm{^{56}Ni}}

\newcommand{\vnimax}{{v_{\rm Ni}^{\rm max}}}

\slugcomment{Accepted by the Astrophysical Journal Letters}

\begin{document}

\title{Nucleosynthesis and Clump Formation in a Core Collapse Supernova}

\author{K. Kifonidis\altaffilmark{1}, 
        T. Plewa\altaffilmark{2,1},
        H.-Th. Janka\altaffilmark{1}
        and E. M\"uller\altaffilmark{1}}
\altaffiltext{1}{Max-Planck-Institut f\"ur Astrophysik 
                 Karl-Schwarzschild-Stra{\ss}e 1, 85740 Garching, 
                 Germany}
\altaffiltext{2}{Nicolaus Copernicus Astronomical Center, 
                 Bartycka 17, 00716 Warsaw, Poland}

\begin{abstract}
High-resolution two-dimensional simulations were performed for the
first five minutes of the evolution of a core collapse supernova
explosion in a 15\,$\Msol$ blue supergiant progenitor.  The
computations start shortly after bounce and include neutrino-matter
interactions by using a light-bulb approximation for the neutrinos,
and a treatment of the nucleosynthesis due to explosive silicon and
oxygen burning.  We find that newly formed iron-group elements are
distributed throughout the inner half of the helium core by
Rayleigh-Taylor instabilities at the Ni+Si/O and C+O/He interfaces,
seeded by convective overturn during the early stages of the
explosion. Fast moving nickel mushrooms with velocities up to $\sim
4000$\,km/s are observed.  This offers a natural explanation for the
mixing required in light curve and spectral synthesis studies of
Type~Ib explosions. A continuation of the calculations to later times,
however, indicates that the iron velocities observed in SN~1987\,A
cannot be reproduced because of a strong deceleration of the clumps in
the dense shell left behind by the shock at the He/H interface.
\end{abstract}

\keywords{hydrodynamics -- instabilities -- nucleosynthesis --
         shock waves -- stars: supernovae}

\section{Introduction}

Overwhelming observational evidence (cf. references in
\citealt{Mueller98}) suggests that large-scale mixing processes took
place in SN~1987\,A and transported newly synthesized $\Ni$ from its
creation site near the collapsed core all the way out to the hydrogen
envelope.  Spectroscopic studies of SN 1987\,F, SN 1988\,A, SN 1993\,J
\citep[][and references therein]{Spyromilio94} and SN 1995\,V
\citep{F+98} indicate that such mixing is probably generic in core
collapse supernovae. Indeed, artificial mixing of the radioactive
ejecta within the (helium) envelope is indispensable in order to
reproduce the light curves and spectra of Type Ib explosions using
one-dimensional (1D) hydrodynamic models (\citealt*{Shigeyama};
\citealt{WE97}, and references therein).

These findings have instigated theoretical work on multidimensional
supernova models focusing either on the role of convective
instabilities in the delayed, neutrino-driven explosion mechanism
within about the first second of evolution (\citealt{Mezz+98};
\citealt{JM96}; \citealt*{BHF95}; \citealt{HBFC94}; \citealt*{Mil93}),
or on the growth of Rayleigh-Taylor instabilitites during the late
evolutionary stages (\citealt*{NSS98}; \citealt{HB92};
\citealt*{MFA91}; \citealt{YS91}; \citealt{HMNS90}).  However,
multidimensional simulations which follow the evolution of the
supernova shock from its revival due to neutrino heating, until its
emergence from the stellar surface have not yet been performed.

In this {\em Letter\/}, we report on preliminary results of
high-resolution two-dimensional (2D) computations which for the first
time cover the neutrino-driven initiation of the explosion, the
accompanying convection and nucleosynthesis as well as the
Rayleigh-Taylor mixing within the first $\sim 300$ seconds of
evolution.

\section{Numerical method and initial data}
\label{sect:numerics}

We split our simulation into two stages.  The early evolution ($t
\lesssim 1$\,s) is followed with a version of the {\sc HERAKLES} code
(T. Plewa \& E. M\"uller, in preparation) which solves the
multidimensional hydrodynamic equations using the direct Eulerian
version of the Piecewise Parabolic Method \citep{CW84} augmented by
the Consistent Multifluid Advection scheme of \cite{CMA} in order to
guarantee exact conservation of nuclear species.  We have added the
input physics described in \cite{JM96} (henceforth JM96) with the
following modifications. General relativistic corrections are made to
the gravitational potential following \cite{Van_Riper}.  A 14-isotope
network is incorporated in order to compute the explosive
nucleosynthesis.  It includes the 13 $\alpha$-nuclei from $\rm ^4He$
to $\Ni$ and a representative tracer nucleus which is used to monitor
the distribution of the neutrino-heated, neutron-rich material and to
replace the $\Ni$ production when $Y_{\rm e}$ drops below $\sim 0.49$
\citep*[cf.][]{TNH96}.  Our initial data are taken from the 15\,$\Msol$
progenitor model of \cite*{WPE88} which was collapsed by \cite{Bruenn}.
The model is mapped to a 2D grid consisting of 400 radial zones ($
3.17\times 10^{6}\,{\rm cm} \leq r \leq 1.7\times10^9$\,{\rm cm}), and
180 angular zones ($0 \leq \theta \leq \pi$; cf. JM96 for details).  A
random initial seed perturbation is added to the velocity field with a
modulus of $10^{-3}$ of the (radial) velocity of the post-collapse
model.  The computations begin 20\,ms after core bounce and are
continued until 885\,ms when the explosion energy has saturated at
$1.48\times10^{51}$\,erg (this value has still to be corrected for the
binding energy of the outer envelope).  We will henceforth refer to
this calculation as our ``explosion model''.

The subsequent shock propagation through the stellar envelope and the
growth of Rayleigh-Taylor instabilities is followed with the AMRA
Adaptive Mesh Refinement (AMR) code (T. Plewa \& E. M\"uller, in
preparation). Neutrino physics and gravity are not included in the AMR
calculations.  Both do not influence the propagation of the shock
during late evolutionary stages.  However, gravity is important for
determining the amount of fallback, a problem which is outside the
scope of the present study.  The equation of state takes into account
contributions from photons, non-degenerate electrons, $\rm
e^+e^-$-pairs, $\rm ^1H$, and the nuclei included in the reaction
network. The AMR calculations are started with the inner and outer
boundaries located at $r_{\rm in}=10^8$\,cm (i.e. inside the hot
bubble containing the neutrino-driven wind) and $r_{\rm
out}=2\times10^{10}$\,cm, respectively.  No further seed perturbations
are added.  Our maximum resolution is equivalent to that of a uniform
grid of $3072 \times 768$ zones.  We do not include the entire star
but expand the radial extent of the grid by a factor of 2 to 4
whenever the supernova shock is approaching the outer grid boundary,
which is moved from its initial value out to $r_{\rm
out}=1.1\times10^{12}$\,cm at $t=300$\,s.  Reflecting boundary
conditions are used at $\theta = 0$ and $\theta = \pi$ and free
outflow is allowed across the inner and outer radial boundaries.

\section{Results}
\label{sect:results}

The general features of our explosion model are comparable to the
models of JM96. The most important difference is caused by our use of
general relativistic corrections to the gravitational potential.
Since the shock has to overcome a deeper potential well than in the
Newtonian case, the luminosities (which are prescribed at the inner
boundary) required to obtain a certain final explosion energy are
roughly 20\% higher than those of JM96.  In particular, we have
adopted the following set of parameters: $L_{\nu_e}^0 = 2.8125 \times
10^{52}\,{\rm erg/s} , L_{\nu_x}^0 = 2.375\times10^{52}\,{\rm erg/s},
\Delta Y_l = 0.0875, \Delta \varepsilon = 0.0625$ (cf. JM96). The
neutrino spectra and the temporal decay of the luminosity are the same
as in JM96.

For the chosen neutrino luminosities the shock starts to move out of
the iron core almost immediately.  Convection between shock and gain
radius sets in $\sim 30$\,ms after the start of the simulation in form
of rising blobs of heated, deleptonized material (with $Y_{\rm e} \ll
0.5$) separated by narrow downflows with $Y_{\rm e} \approx 0.49$.
The shock reaches the Fe/Si interface at $r = 1.4\times10^8$\,cm after
$\sim 100$\,ms. At this time the shocked material is still in nuclear
statistical equilibrium and is composed mainly of $\alpha$-particles
and nucleons. When the temperature right behind the shock drops below $\sim 7
\times 10^9$\,K, $\Ni$ starts to form in a narrow shell.  During the
ongoing expansion and cooling, nickel is also synthesized in the
convective region. However, this synthesis proceeds exclusively in the
narrow downflows which separate the rising bubbles and have a
sufficiently high electron fraction $Y_{\rm e}$.  The convective
anisotropies therefore lead to a highly inhomogeneous nickel
distribution (Fig.~\ref{fig:hotb}). 

Freezeout of complete silicon
burning occurs at $t \approx 250$\,ms, and convection ceases at $t
\approx 400$\,ms, at which time the flow pattern becomes frozen in.
Subsequently, the entire post-shock region expands nearly
uniformly. The post-shock temperature drops below $2.8 \times 10^9$\,K
at $t = 495$\,ms, when the shock is about to cross the Si/O
interface. Thus, our model shows only moderate oxygen burning (due to
a non-vanishing oxygen abundance in the silicon shell), and negligible
neon and carbon burning.  This is caused by the specific structure of
the progenitor model of \cite{WPE88} and may change when different
(e.g.\ more massive) progenitors are used.  In total, $0.052\,\Msol$
of $\Ni$ are produced, while $0.10\,\Msol$ of material are synthesized
at conditions with $Y_{\rm e} < 0.49$ and end up as neutron-rich
nuclei.  Fig.~\ref{fig:hotb} shows the $\Ni$ distribution at $t=
885$\,ms.  The density ratio between the dense regions which contain
the nickel and the low-density, deleptonized material in the bubbles
deeper inside is $\sim 2.5$.

During the next seconds the shock detaches from the formerly
convective shell that carries the products of explosive
nucleosynthesis and crosses the C+O/He-interface, initially
accelerating and subsequently decelerating due to the varying density
gradient.  Twenty seconds after core bounce this unsteady propagation
speed of the shock has led to a strong compression of the
metal-containing shell. At the inner boundary of the shell pressure
waves have steepened into a reverse shock.  Rayleigh-Taylor
instabilities start to grow at the Ni+Si/O- and C+O/He-interfaces and
are fully developed and interact with each other at $t=50$\,s. Nickel
and silicon are dragged upward into the helium shell in rising
mushrooms on angular scales from $1^{\circ}$ to about $5^{\circ}$,
whereas helium is mixed inward in bubbles. Oxygen and carbon, located
in intermediate layers of the progenitor, are swept outward as well as
inward in rising and sinking flows.
At $t = 300$\,s the densities and radial velocities between the dense
mushrooms and clumps and the ambient medium differ by factors up to 5
and 1.3, respectively. As Fig.~\ref{fig:AMRA} shows, the fastest
mushrooms have already propagated out to more than half the radius of
the He core. The excessive outflows along the symmetry axis in
Fig.~\ref{fig:AMRA} are a numerical artifact caused by the singularity
of the polar axis in spherical coordinates. In order not to introduce
large errors, only an angular wedge with $15^{\circ} \leq \theta \leq
165^{\circ}$ was used in all the analyses.

The remarkable efficiency of the mixing is clearly visible from
Fig.~\ref{fig:mixing} where the original onion-shell structure of the
progenitor's metal core has disappeared after 300\,s.  We observe
large composition gradients between different regions of a mushroom
and also between different mushrooms. Some of them contain $\Ni$ mass
fractions of more than 70\% whereas others have only nickel admixtures
of 20\% or less but show high concentrations of silicon and
oxygen. Composition contrasts of at least this magnitude have recently
been observed in different filaments of the Cas~A supernova remnant by
\cite{Hughes99}. We find that also more than $0.04\,\Msol$ of matter
consisting of neutron-rich nuclei are mixed into the $\Ni$ and $\rm
^{28}Si$ clumps.

The reverse shock substantially decelerates the bulk of the nickel
from $\sim 15\,000$\,km/s at $t=885$\,ms to 3200 -- 4500\,km/s at
$t=50$\,s, at which time the maximum velocities $\vnimax$ have dropped
to 5800\,km/s. After another 50\,s, however, the clumps start to move
essentially ballistically through the helium core and only a slight
drop of $\vnimax$ from $5000$~km/s at $t=100$\,s to $4700$\,km/s at
$t=300$\,s occurs, when most of the $\Ni$ has velocities below
$3000$\,km/s (Fig.\ref{fig:nickelv}).

We have recently accomplished to follow the subsequent evolution of
our model up to 16\,000\,s after core bounce. Similar to the situation
at the C+O/He interface, the supernova shock leaves behind a dense
(Rayleigh-Taylor unstable) shell at the He/H interface.  While the
entire shell is rapidly slowed down, a second reverse shock forms at
its inner boundary (Fig.~\ref{fig:AMRA}). Our high-resolution
simulations reveal a potentially severe problem for the mixing of
heavy elements into the hydrogen envelope of Type II supernovae like
SN~1987A. We find that the fast nickel containing clumps, after having
penetrated through this reverse shock, dissipate a large fraction of
their kinetic energy in bow shocks created by their supersonic motion
through the shell medium. This leads to their deceleration to $\sim
2000$~km/s in our calculations, a negative effect on the clump
propagation which has not been discussed previously. A detailed
analysis of this result will be subject of a forthcoming publication.

\section{Conclusions}
\label{sect:conlusions}

Our calculations provide a {\em natural\/} explanation for the mixing
of $\Ni$ into the helium layer, which is required to reproduce the
light curves of Type Ib supernovae \citep{Shigeyama}. In this sense
they justify the rather large seed perturbations ($\geq 5\%$) which
were imposed by \cite{HMNS94} on the radial velocity given from
spherically symmetric models of exploding helium cores at about
10$\,$s after shock formation. In addition, they may have interesting
implications for the modeling of Type Ib spectra (Woosley, private
communication; compare Fig.~\ref{fig:mixing} of this work with Fig.~7
in \citealt{WE97}).

Our simulations suggest that ballistically moving, metal-rich clumps
with velocities up to more than $\sim 4000$\,km/s are ejected during
the explosion of Type Ib (and Ic) supernovae. In case of Type II
supernovae, however, the dense shell left behind by the shock passing
the boundary between helium core and hydrogen envelope, causes a
substantial deceleration of the clumps. The high iron velocities
observed in SN~1987\,A, for example, can therefore not be accounted
for by our models. This problem would be reduced if the density
profile at the He/H interface were smoother, leading to less strong
variations of the shock velocity. Could this direct to a possible
common envelope phase or merger history of the progenitor star
(\citealt{HM89}; \citealt*{PJR90})?  Alternatively, the high iron
velocities in SN~1987\,A could require additional energy input from Ni
decay \citep{HB92} or could imply a large global anisotropy of the
explosion, e.g. associated with jets emerging from the collapse of a
rapidly rotating stellar core \citep{FH99}. Though final conclusions
require three-dimensional calculations, our simulations indicate that
all computations of Rayleigh-Taylor mixing in Type~II supernovae
carried out so far (including the case of SN~1987\,A; \citealt{HB92})
have been started from overly simplified initial conditions since they
have neglected clump formation within the first minutes of the
explosion.

The mixing of nucleosynthesis products as seen in our models poses
another potential problem. A significant fraction ($0.04\,\Msol$ of a
total of $0.1\,\Msol$) of the neutron-rich nuclei which are
synthesized in regions with $Y_{\rm e}< 0.49$ is dragged outward by
the nickel containing mushrooms and clumps. At least this amount will
most likely be ejected in the explosion. A detailed analysis of the
nuclear composition is necessary to tell whether this is in conflict
with limits from Galactic chemical evolution models, which allow for
at most $10^{-3}$--$10^{-2}\,\Msol$ of neutron-rich nuclei to be
ejected per supernova event (e.g., \citealt{HBFC94}; \citealt{TNH96}).
Later fallback will not solve this potential problem: How could it
disentangle the clumpy nickel ejecta from their undesirable pollution?
A better knowledge of the luminosities and spectra of electron
neutrinos and antineutrinos emitted from the nascent neutron star is
therefore needed to perform more reliable calculations of the
neutronization of the neutrino-heated ejecta.

During the computations we became aware of oscillations with angle in
parts of the postshock flow (Figs.~\ref{fig:hotb} and
\ref{fig:AMRA}). These are caused by the ``odd-even decoupling''
phenomenon associated with grid-aligned shocks \citep{JJQ}.  As a
consequence, the maximum nickel velocities, $\vnimax$, obtained in our
AMR calculations have probably been overestimated by $\sim 25\%$
because the growth of some of the mushrooms was influenced by the
perturbations induced by this numerical defect. The main results of
our study, however, are not affected. We note that a large number of
supernova calculations performed with codes based either on the direct
Eulerian (cf.  Figs.~22 and 24 in \citealt{BHF95}, Fig.~20 in JM96),
or the Lagrangean with remap formulation of the PPM scheme
\citep{Mezz+98} seem to be affected by this numerical flaw.  We defer
a detailed analysis of this problem to a forthcoming publication.

\acknowledgments

We thank S. Woosley for profiles of the progenitor star and S. Bruenn
for his post-bounce core model. We acknowledge support by P. Cieciel\c
ag and R. Walder concerning visualization and by the crew of the
Rechenzentrum Garching where the simulations were performed on the NEC
SX-4B and CRAY J916 computers. TP was supported by grant 2.P03D.004.13
from the Polish Committee for Scientific Research, HTJ by DFG grant
SFB-375 f\"ur Astro-Teilchenphysik.

\clearpage

\begin{figure}
\caption{
Distribution of the $\Ni$ mass fraction at $t=885$\,ms.
                      At this time the shock (not visible) has almost
                      reached the outer boundary of the computational
                      domain (blue) at $\rm 1.7\times10^9~cm$.}
\label{fig:hotb}
\end{figure}

\clearpage

\begin{figure}
\caption{Density distribution ($\log_{10} \rho~{\rm [g\,cm^{-3}]}$ ) 300\,s  after
         core bounce in the inner $\sim 3\times10^{11}$\,cm of the star.
         The supernova shock (outermost blue
         discontinuity) is located inside the
         hydrogen envelope at $r = 2.7\times10^{11}$\,cm. 
         A dense shell (visible as a red ring) 
         has formed behind the shock. Its outer boundary coincides with 
         the He/H interface while its inner boundary is in the process
         of steepening into a reverse shock.}
\label{fig:AMRA}
\end{figure}

\clearpage

\begin{figure}
\begin{center}
\plotone{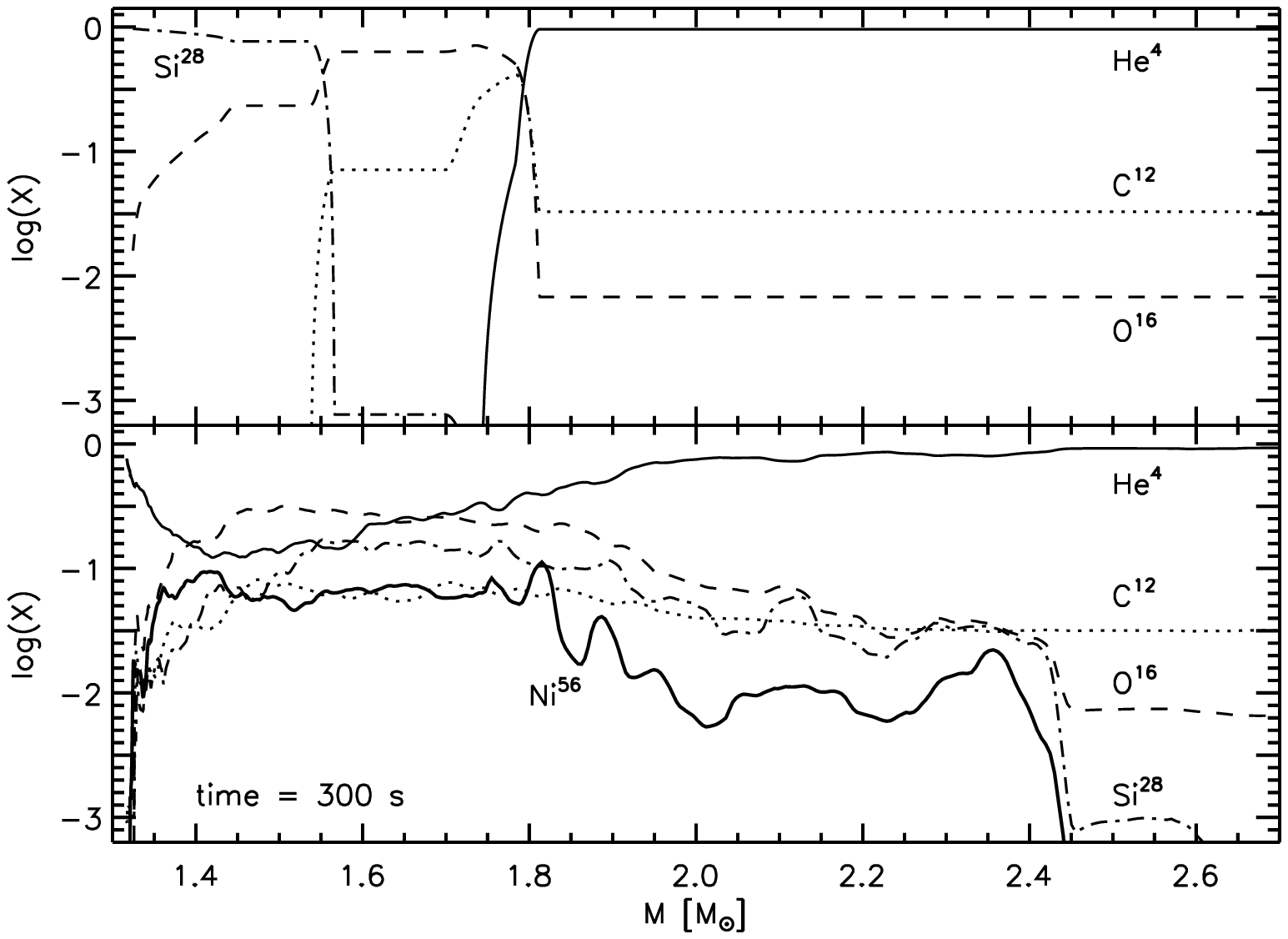}
\end{center}
\caption{Initial composition exterior to the iron core (top)
         and composition 300~s after core bounce (bottom). The
         C/O core of the progenitor has been completely shredded by the
         Rayleigh-Taylor instability.
         $\rm ^{12}C$, $\rm ^{16}O$, $\rm ^{28}Si$ and the newly
         synthesized  $\Ni$ have been mixed beyond the inner half of
         the helium core ($M_{\rm He-core} = 4.2\,\Msol$) to a mass
         coordinate of $2.45\,\Msol$. Inward mixing of $\rm ^{4}He$ 
         has led to an increased helium concentration interior to 
         1.75\,$\Msol$. Inside $1.4\,\Msol$ the $\alpha$-rich freezeout
         in the neutrino-heated ejecta yields a contribution of $^4$He.
         A cone with opening angle of 15 degrees around the polar axis
         was excluded from the analysis (see text).}
\label{fig:mixing}
\end{figure}

\clearpage

\begin{figure}
\begin{center}
\plotone{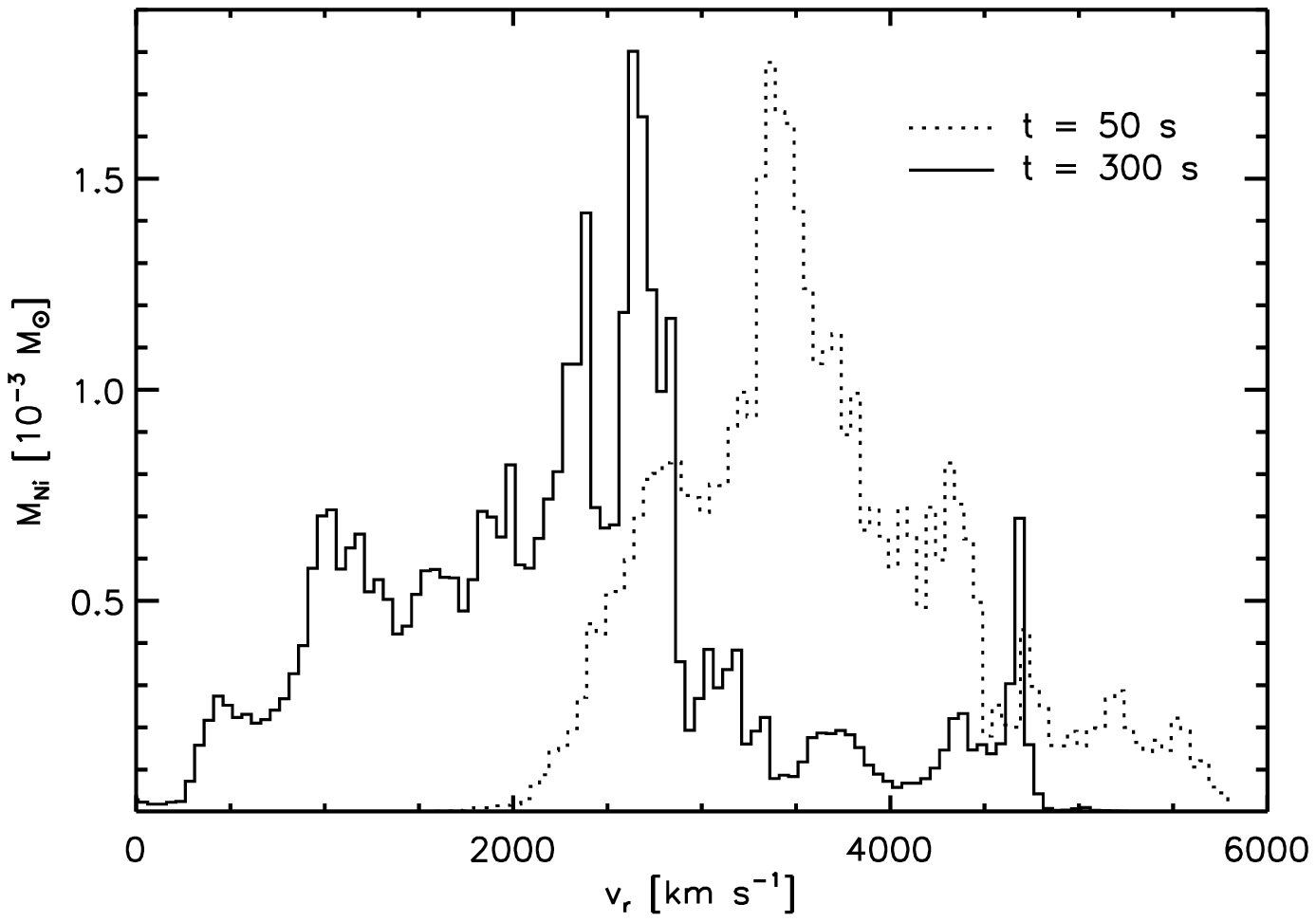}
\end{center}
\caption{Distribution of the $\Ni$-mass versus velocity at $t=50$\,s
         and $t=300$\,s, respectively. A cone with opening angle of 15 degrees 
         around the polar axis was excluded from the analysis 
         (see text).}
\label{fig:nickelv}
\end{figure}


\begin{thebibliography}{}

\bibitem[Bruenn(1993)]{Bruenn}  Bruenn, S. W.  1993, in Nuclear Physics
    in the Universe, ed. M. W. Guidry \& M. R. Strayer, (Bristol: IOP), 31

\bibitem[Burrows et al.(1995)Burrows, Hayes, \& Fryxell]{BHF95} Burrows, A.,
    Hayes, J., \& Fryxell, B. A. 1995, \apj, 450, 830

\bibitem[Colella \& Woodward(1984)]{CW84} Colella, P.,
    \& Woodward, P. R. 1984, J. Comput. Phys., 54, 174

\bibitem[Fassia et al.(1998)]{F+98} Fassia, A., Meikle, W. P. S., 
             Geballe, T. R., Walton, N. A., Pollacco, D. L.,
             Rutten, R. G. M., \& Tinney, C. 1998, \mnras, 299, 150

\bibitem[Fryer \& Heger(1999)]{FH99} Fryer, C. L., \& Heger, A. 1999,
      \apj, submitted (astro-ph/9907433)

\bibitem[Hachisu et al.(1990)]{HMNS90}  Hachisu, I.,
     Matsuda, T., Nomoto, K., \& Shigeyama, T. 1990, \apjl, 358, L57

\bibitem[Hachisu et al.(1994)]{HMNS94}  
     Hachisu, I., Matsuda, T., Nomoto, K., \& Shigeyama, T. 1994, \aaps, 104, 341

\bibitem[Herant \& Benz(1992)]{HB92} Herant, M.,
    \& Benz, W.  1992, \apj, 387, 294

\bibitem[Herant et al.(1994)]{HBFC94} Herant, M.,
     Benz, W., Hix, W. R., Fryer, C. L., \& Colgate, S. A. 1994, \apj, 435, 339

\bibitem[Hillebrandt \& Meyer(1989)]{HM89} Hillebrandt, W., \&
    Meyer, F. 1989, \aap, 219, L3

\bibitem[Hughes et al.(2000)]{Hughes99} Hughes, J. P.,
    Rakowski, C. E., Burrows, D. N., \& Slane, P. O. 2000, \apjl,
    528, L109

\bibitem[Janka \& M\"uller(1996)]{JM96} Janka, H.-Th., \& M\"uller, E. 1996, 
    \aap, 306, 167 (JM96)

\bibitem[Mezzacappa et al.(1998)]{Mezz+98} Mezzacappa, A., 
     Calder, A. C., Bruenn, S. W., Blondin, J. M.,
     Guidry, M. W., Strayer, M. R., \& Umar,  A. S. 1998, \apj, 495, 911           

\bibitem[Miller et al.(1993)Miller, Wilson, \& Mayle]{Mil93} Miller, D. S.,
    Wilson, J. R., \& Mayle, R. W.  1993, \apj, 415, 278

\bibitem[M\"uller(1998)]{Mueller98}  M\"uller, E.  1998, in 
        Computational Methods for Astrophysical Fluid Flow, ed.
        O. Steiner \& A. Gautschy, (Berlin: Springer), 371

\bibitem[M\"uller et al.(1991)M\"uller, Fryxell, \& Arnett]{MFA91} M\"uller, E.,
    Fryxell, B. A., \& Arnett, W. D.  1991, \aap, 251, 505

\bibitem[Nagataki et al.(1998)Nagataki, Shimizu, \& Sato]{NSS98} Nagataki, S.,
    Shimizu, T. M., \& Sato, K.  1998, \apj, 495, 413

\bibitem[Plewa \& M\"uller(1999)]{CMA} Plewa, T., \& M\"uller, E. 1999, \aap,
    342, 179

\bibitem[Podsiadlowski et al.(1990)Podsiadlowski, Joss, \& Rappaport]{PJR90} Podsiadlowski, P., 
    Joss, P. C., \& Rappaport, S. 1990, \aap, 227, L9

\bibitem[Quirk(1994)]{JJQ} Quirk, J. J. 1994,
    Int. J. Num. Meth. Fluids, 18, 555

\bibitem[Shigeyama et al.(1990)Shigeyama,
          Nomoto, \& Tsujimoto]{Shigeyama} Shigeyama, T.,
          Nomoto, K., \& Tsujimoto T. 1990, \apjl, 361, L23

\bibitem[Spyromilio(1994)]{Spyromilio94} Spyromilio, J. 1994, \mnras, 266, L61

\bibitem[Thielemann et al.(1996)Thielemann,
    Nomoto, \& Hashimoto]{TNH96} Thielemann, F.-K., 
    Nomoto, K., \& Hashimoto, M.  1996, \apj, 460, 408

\bibitem[Van Riper(1979)]{Van_Riper} Van Riper, K. A. 1979, \apj, 232, 558

\bibitem[Woosley et al.(1988)Woosley,
    Pinto, \& Ensman]{WPE88} Woosley, S. E.,
    Pinto, P. A., \& Ensman, L.  1988, \apj, 324, 466

\bibitem[Woosley \& Eastman(1997)]{WE97} Woosley, S. E.,
    \& Eastman, R.  1997, in Thermonuclear Supernovae, ed.
    P. Ruiz-LaPuente, R. Canal, \& J. Isern, (Dordrecht: Kluwer), 821

\bibitem[Yamada \& Sato(1991)]{YS91} Yamada, S.,
    \& Sato, K.  1991, \apj, 382, 594


\end{thebibliography}
\end{document}